\documentclass[%
 reprint,
 amsmath,amssymb,
 aps,
]{revtex4-1}

\usepackage{graphicx}% Include figure files
\usepackage{dcolumn}% Align table columns on decimal point
\usepackage{bm}% bold math
\begin{document}

\preprint{APS/123-QED}
\begin{flushright}
IZTECH-2015-03
\end{flushright}

\title{Separate Einstein-Eddington Spaces and the Cosmological Constant}
\author{Hemza Azri}
 \email{Email: hemzaazri@iyte.edu.tr}
\affiliation{%
 Department of Physics, {\.I}zmir Institute of Technology, TR35430, {\.I}zmir, Turkey
}%

\begin{abstract}
Based on Eddington affine variational principle on a locally product manifold, we derive the separate Einstein space described by its Ricci tensor. The derived field equations split into two field equations of motion that describe two maximally symmetric spaces with two cosmological constants. We argue that the invariance of the {\it bi-field equations} under {\it projections} on the separate spaces, may render {\it one} of the cosmological constants to zero. We also formulate the model in the presence of a  scalar field. The resulted separate Einstein-Eddington spaces maybe considered as two states that describe the universe before and after inflation. A possibly interesting affine action for a general perfect fluid is also proposed. It turns out that the condition which leads to zero cosmological constant in the vacuum case, eliminates here the effects of the gravitational mass density of the perfect fluid, and the dynamic of the universe in its final state is governed by only the inertial mass density of the fluid.
\begin{description}
\item[Keywords]
Purely affine gravity, Eddington's gravity, Separate Einstein spaces, product spaces, cosmological constant, vacuum energy, inflation.
\keywords{Suggested keywords}
\end{description}
\end{abstract}
\maketitle

%\tableofcontents

\section{Introduction}

In the purely metric formulation of gravity based on Einstein-Hilbert action, no fundamental symmetry principle explains the existence of the metric tensor, it is assumed a priori in the spacetime. Attempts have been made to study gravitational interactions as in the framework of Yang Mills theories where the fundamental fields are explained by some fundamental symmetry principles (local gauge invariance). This leads to postulating affine Lagrangians which are of first order of a GL$\left(4, R\right)$ affine connection $\Gamma$.

Although it seems reasonable postulating affine action principle, defining an invariant volume element in a background without a metric tensor is a serious problem. The simplest volume element that can be proposed is a density constructed from the Ricci tensor, which is of course a linear function of the connection. This was proposed by Eddington and it became the first and the simplest affine theory of gravity in a background free of any matter fields \cite{edd1,edd2}. In this theory, the metric tensor is generated from a dynamical equations derived from an affine action principle.

Purely affine formulation of gravity in the presence of certain matter fields has been discussed in details in \cite{kijowski1,kijowski2}. In this structure, the metric tensor $g$ which is not referred a priori in the action, is taken as a canonically conjugate to the affine connection $\Gamma$. This picture is found to be more fundamental than purely metric theory in quantum gravity. In fact, a theory of gravity written only in terms of a non metric affine connection is found to be a unitary, power counting renormalizable \cite{martellini1,martellini2}.

An interesting theory of gravity based on both metric and affine connection has been proposed where the two quantities are independent \cite{banados}.

Generalizing affine theory of  gravity to nonsymmetric connection, i.e, in the presence of torsion tensor, has been discussed in \cite{poplawski1,poplawski2}.

It was also shown that matter terms can be generated dynamically when extending Eddington's gravity with Riemann tensor \cite{demir1}.

Recently, Eddington's gravity has been formulated in a four dimensional spacetime that was considered plunged in a larger eight dimensional space, and it was shown that matter terms could appear in the new structure \cite{azri1}.

In this paper, we derive the separate Einstein spaces from affine theory. These spaces are studied in details in Riemannian geometry, they are locally product spaces endowed with a metric structure, where their two components have constant curvature. Without postulating a metric structure a priori, we will derive these spaces from an action principle where the field configuration is an affine connection. The resulting equations describe two universes dominated by two cosmological constants.

We will tackle the cosmological constant problem in this formalism and show that under some symmetry, which is proposed to be the invariance under the projection on the separate spaces, one of the cosmological constants automatically vanishes. We will also formulate the model in the presence of matter fields. In this case, we discuss the possibility that the resulted spaces (described by two field equations) may be considered as two states which describe the universe before and after inflation.

In cosmology, matter fields are usually supposed to be perfect fluids with different equations of states. In general relativity the form of these fluids are generally postulated. However, we find it interesting here to propose an affine action for these fluids. In addition to the affine connection  and the directional four vector velocities, this action is based on the so called \text{\it inertial} mass density. This mass density appears explicitly in the action and it plays an important role in its definition. We will see that the condition which zeroed the cosmological constant in the vacuum case will only eliminate the effects of the gravitational mass density of the perfect fluid in this case.

The paper is organized as follows, in section two we propose an affine Lagrangian density based on an affine connection in a $2N$ dimensional space, and derive the separate Einstein equations from variational principle. In section three, we discuss how cosmological constant vanishes automatically in one of the spaces due to the invariance under projection, and we formulate the model in the presence of matter. In section four we summarize.

\section{Separate Einstein-Eddington space}

In Einstein-Hilbert action principle, the spacetime is supposed to be pseudo-Riemannian space, i.e, a space endowed with a metric tensor (with a signature) which describes the intervals (distances and times) between different events in the curved background. This tensor field defines completely the Levi-Civita connection which in turn defines the associated Riemann tensor.

However, defining the Riemann tensor does not provide a metric field. In fact, that is the rule of parallel displacement which provides a measure of the curvature of the manifolds. This rule is incorporated in the so called affine connection via covariant derivatives. Theories of gravity which are based on this affine connections as fundamental fields are called purely affine theories.

The simplest affine theory of gravity is derived from Eddington lagrangian density which is defined by the square root of the determinant of the Ricci tensor \cite{edd1,edd2}. Like Einstein's general theory of relativity, Eddington's gravity can be extended to higher dimensions \cite{azri1}.

In this section, we will formulate Eddington's affine action principle in a higher dimensional space which is supposed to have a product structure. The reason of taking the product structure is to derive the separate Einstein's space for which we will give a brief definition in what follows.

We suppose that we are given a $2N$-dimensional space which admits a locally product structure, this is defined by the existence of a separating coordinate system $x^{j}$ such that in any intersection of two neighbourhoods $x^{k}$ and $x^{k\prime}$ one has \cite{tachibana,yano}
\begin{equation}
x^{\mu\prime}=x^{\mu\prime}\left(x^{\mu} \right), \quad \quad
x^{\mu^{\ast}\prime}=x^{\mu^{\ast}\prime}\left(x^{\mu^{\ast}} \right),
\end{equation}
where $\mu=1,...,N$ and $\mu^{\ast}=N+1,...,2N$.
\newline
In other words, the global space appears as the product of two spaces $\mathcal{M}$ and $\mathcal{M}^{\ast}$ defined by their coordinate systems $x^{\mu}$ and $x^{\mu^{\ast}}$ respectively.

If this space is endowed with a metric a priori, the separate Einstein's spaces are product spaces which described by their ricci tensors $\mathcal{R}_{ij}$ that split into its components as follows \cite{tachibana,yano}
\begin{equation}
\mathcal{R}_{\mu\nu}= (a+b) g_{\mu\nu},\quad \mathcal{R}_{\mu^{\ast}\nu^{\ast}}= (a-b) g_{\mu^{\ast}\nu^{\ast}},
\end{equation}
where $a$ and $b$ are constants.
\newline
These equations describe two maximally symmetric spaces, i.e, spaces of constant curvatures. These curvatures are given by the two non independent cosmological terms $a+b$ and $a-b$ respectively.

In what follows, we will derive these field equations from affine variational principle where the metric tensor appears from dynamical equations without postulating it a priori.
\newline
To do this, let us suppose that the $2N$-dimensional product space is endowed with an affine connection given by its components $\Gamma_{ij}^{k}$, where $i,j=1,...2N$.
\newline
The curvature tensor with components $\mathcal{R}_{ijk}^{l}$ is defined in usual way with respect to the affine connection as
\begin{equation}
\mathcal{R}_{ijk}^{l}=\partial_{i}\Gamma_{jk}^{l}-\partial_{j}\Gamma_{ik}^{l}+\Gamma_{im}^{l}\Gamma_{jk}^{m}-\Gamma_{jm}^{l}\Gamma_{ik}^{m}.
\end{equation}
By definition, the Ricci tensor $\mathcal{R}_{ij}$ is given by
\begin{equation}
\mathcal{R}_{ij}=\mathcal{R}_{ikj}^{k}.
\end{equation}
In this space, Eddington's gravity is based on the following action
\begin{equation}
S= 2 \int d^{2N} x \sqrt{\text{Det}\left[ \mathcal{R}_{ij}\right]},
\end{equation}
where we took Eddington's Lagrangian density
\begin{equation}
\label{lagran1}
\mathcal{L}= 2\sqrt{\text{Det}\left[ \mathcal{R}_{ij}\right]},
\end{equation}
with $\text{Det}\left[ \mathcal{R}_{ij}\right]$ is the determinant of the symmetric part of the Ricci tensor, and the connection $\Gamma$ is supposed to be symmetric (torsionless).
\newline
Once we defined a Lagrangian density in which the field configuration is the connection $\Gamma$ which is embodied in the Ricci tensor $\mathcal{R}_{ij}$, a canonical momentum conjugate to $\Gamma$ can be defined as \cite{kijowski1,kijowski2}
\begin{eqnarray}
\label{pi}
\pi^{ij}= \frac{\partial \mathcal{L}}{\partial \mathcal{R}_{ij}}.
\end{eqnarray}
This new scalar density, is the origin of the metric structure in the theory as we shall see later.
With the Lagrangian density (\ref{lagran1}), equation (\ref{pi}) is equivalent to
\begin{equation}
\label{pi1}
\sqrt{\text{Det}\left[ \mathcal{R}_{ij}\right]}\mathcal{R}^{ij}=\pi^{ij},
\end{equation}
where $\mathcal{R}^{ij}$ is the inverse of the Ricci tensor defined such that $\mathcal{R}^{ik}\mathcal{R}_{kj}=\delta^{i}_{j}$.
\newline
As in field theory, the dynamical equations are derived from the following local Euler-Lagrange equations
\begin{equation}
\label{euler}
\partial_{l}\left( \frac{\partial\mathcal{L}}{\partial\left(  \partial_{l}\Gamma^{i}_{jk}\right)} \right) - \frac{\partial\mathcal{L}}{\partial\Gamma^{i}_{jk}}=0.
\end{equation}
When applied to the Lagrangian density (\ref{lagran1}) with the use of relation (\ref{pi}), the last equations give the dynamical equation
\begin{equation}
\label{dyn}
\nabla_{k} \pi^{ij}=0.
\end{equation}
Here, the covariant derivative operator $\nabla$ is with respect to the affine connection $\Gamma$.

One may propose a solution to the dynamical equation (\ref{dyn}) by introducing an invertible, covariantly constant and symmetric tensor field $\mathcal{G}_{ij}$ such that
\begin{equation}
\label{metric}
\pi^{ij}=\sqrt{\text{Det}\left[\mathcal{G}_{ij}\right]}\mathcal{G}^{ij}.
\end{equation}
In general, the tensor field $\mathcal{G}_{ij}$ is given by its components
\begin{equation}
g_{\mu\nu}, \quad g_{\mu\nu^{\ast}}, \quad g_{\mu^{\ast}\nu}, \quad g_{\mu^{\ast}\nu^{\ast}}.
\end{equation}
We choose the tensor field $\mathcal{G}_{ij}$ given in (\ref{metric}) to be pure, i.e,
\begin{equation}
\mathcal{G}_{ij}=
\left( {\begin{array}{cc}
g_{\mu\nu} & 0 \\
0 & g_{\mu^{\ast}\nu^{\ast}}
\end{array} } \right).
\end{equation}
However, not only $\mathcal{G}_{ij}$ which satisfies $\nabla_{k}\mathcal{G}_{ij}=0$, indeed one may define the pure tensor field $\mathcal{F}_{ij}$ as
\begin{equation}
\label{f}
\mathcal{F}_{ij}=
\left( {\begin{array}{cc}
g_{\mu\nu} & 0 \\
0 & -g_{\mu^{\ast}\nu^{\ast}}
\end{array} } \right).
\end{equation}
The form of the tensor field $\mathcal{F}_{ij}$ given in (\ref{f}) may appear non trivial choice, but as we shall see later, this tensor is important in the product spaces, it is from this tensor field that we can define projective operators which map the global space on $\mathcal{M}$ and $\mathcal{M}^{\ast}$.

This allows us to put a general solution to the dynamical equation (\ref{dyn}) as
\begin{equation}
\label{sol1}
\pi^{ij}=\sqrt{\text{Det}\left[a\mathcal{G}_{ij}+b\mathcal{F}_{ij} \right]} \left( a\mathcal{G}+b\mathcal{F}\right)^{ij},
\end{equation}
where $a, b$ are constants, and $\left( a\mathcal{G}+b\mathcal{F}\right)^{ij}$ is the inverse of the tensor $a\mathcal{G}_{ij}+b\mathcal{F}_{ij}$.
\newline
Equations (\ref{pi1}) and (\ref{sol1}) are equivalent to the following field equations in $2N$ dimension
\begin{equation}
\label{field1}
\mathcal{R}_{ij}=\left( a\mathcal{G}_{ij}+b\mathcal{F}_{ij} \right).
\end{equation}
Now equation (\ref{dyn}) spites into two dynamical equations in both spaces $\mathcal{M}$ and $\mathcal{M^{\ast}}$ respectively as
\begin{equation}
\label{dyn2}
\nabla_{\kappa}g_{\mu\nu}=0, \quad \text{and} \quad \nabla_{\kappa^{\ast}}g_{\mu^{\ast}\nu^{\ast}}=0.
\end{equation}
These dynamical equations generate a metric structure in both spaces, where the resulted metric tensors $g_{\mu\nu}$ and $g_{\mu^{\ast}\nu^{\ast}}$ define a two separate Levi-Civita connections
\begin{equation}
\Gamma^{\mu}_{\alpha\beta}= \frac{1}{2} g^{\mu\lambda} \left( \partial_{\alpha} g_{\beta\lambda} +
\partial_{\beta} g_{\lambda\alpha} - \partial_{\lambda} g_{\alpha\beta}\right)
\end{equation}
\begin{equation}
\Gamma^{\mu^{\ast}}_{\alpha^{\ast}\beta^{\ast}}= \frac{1}{2} g^{\mu^{\ast}\lambda^{\ast}} \left( \partial_{\alpha^{\ast}} g_{\beta^{\ast}\lambda^{\ast}} +
\partial_{\beta^{\ast}} g_{\lambda^{\ast}\alpha^{\ast}} - \partial_{\lambda^{\ast}} g_{\alpha^{\ast}\beta^{\ast}}\right)
\end{equation}
in both spaces $\mathcal{M}$ and $\mathcal{M}^{\ast}$ respectively.

Once the metric structure is generated, one may define a mapping, which projects every vector $v^{i}$ that can be defined by its components $\left(v^{\mu}, v^{\mu^{\ast}} \right)$ in the global space, onto the spaces $\mathcal{M}$ and $\mathcal{M}^{\ast}$. This can be seen by defining the following projection operators \cite{tachibana, yano}
\begin{equation}
\label{proj}
\mathcal{P}_{ij}= \frac{1}{2} \left( \mathcal{G}_{ij}+\mathcal{F}_{ij}\right), \quad \text{and} \quad \mathcal{Q}_{ij}= \frac{1}{2} \left( \mathcal{G}_{ij}-\mathcal{F}_{ij}\right),
\end{equation}
and then
\begin{equation}
\mathcal{P}_{i}^{k}v^{i}=\left(v^{\mu},0 \right), \quad \text{and} \quad
\mathcal{Q}_{i}^{k}v^{i}=\left(0,v^{\mu^{\ast}} \right),
\end{equation}
where $\mathcal{P}_{i}^{k}=\mathcal{G}^{kl} \mathcal{P}_{li}$ and $\mathcal{Q}_{i}^{k}=\mathcal{G}^{kl} \mathcal{Q}_{li}$.

Due to the separability of the space, equation (\ref{field1}) splits into two field equations in the $N$-dimensional spaces $\mathcal{M}$ and $\mathcal{M}^{\ast}$ respectively as follows
\begin{eqnarray}
\label{field2}
\mathcal{R}_{\mu\nu}=\left( a+b\right) g_{\mu\nu},
\end{eqnarray}
and
\begin{eqnarray}
\label{field3}
\mathcal{R}_{\mu^{\ast}\nu^{\ast}}=\left( a-b\right)g_{\mu^{\ast}\nu^{\ast}}.
\end{eqnarray}

The space described by the Ricci tensor (\ref{field1}) is called a separate Einstein space. Equations (\ref{field1}), (\ref{field2}) and (\ref{field3}) are studied in details in the so called locally decomposable Riemannian spaces where $\mathcal{G}_{ij}$ is envisaged as the Riemannian metric given in the space \cite{tachibana,yano}.

Without notion of metric, the derivation given here is completely different, it is based on a covariant Lagrangian density where the fundamental quantity is the affine connection. In the following section we will study some specific symmetry of these spaces and its relation to the cosmological constant problem.

\section{Zero cosmological constant from {\it projective} symmetry}

In cosmology, vacuum energy and the cosmological constant are usually thought to be the same physical quantity. In fact, both quantities appear in Einstein's field equations as source described by an energy momentum tensor of the form $\Lambda_{\text{eff}}g_{\mu\nu}$, where $\Lambda_{\text{eff}}$ can include both terms.

Cosmological observations provide an upper bound of the vacuum energy density of the order $10^{-47}$ $\text{GeV}^{4}$ \cite{planck}. However, theoretical estimates of the ground states of particle fields turned out to be of the order $\Lambda^{4}_{\text{UV}}$, where $\Lambda_{\text{UV}}$ is the momentum Cutoff. In the Planck scale, $\Lambda_{\text{UV}} \sim \text{ M}_{\text{Pl}}$, this is about $10^{76}$  $\text{GeV}^{4}$ which is about a $120$ orders of magnitude larger than the mentioned observed value. This contradiction between theory and observation is the origin of the cosmological constant problem \cite{weinberg,sahni}.

In this section we discuss a way to render the cosmological constant to zero by considering the separate Einstein spaces discussed earlier. We will take the case in which the higher dimensional product space discussed in the previous section is a product of four dimensional spaces, i.e, $N=8$ \cite{azri1,azri2}.

In the previous section, we derived two maximally symmetric universes with two non independent cosmological constants
\begin{equation}
\Lambda=a+b \quad \text{and} \quad \Lambda^{\ast}=a-b,
\end{equation}
where $a$ and $b$ are constants.
\newline
As we see, these constants can never vanish simultaneously unless $a=b=0$.
\newline
Nevertheless, one of the universes can be free of the cosmological constant once the other one takes a specific constant curvature. In fact, for $b=-a$, the field equations (\ref{field2}) and (\ref{field3}) become
\begin{equation}
\label{bi1}
\mathcal{R}_{\mu\nu}=0, \quad
\mathcal{R}_{\mu^{\ast}\nu^{\ast}}=2a g_{\mu^{\ast}\nu^{\ast}}.
\end{equation}
In this case, the space $\mathcal{M}$ is free of the cosmological constant. In the same way, the case $b=a$ gives
\begin{equation}
\mathcal{R}_{\mu\nu}=2a g_{\mu\nu} \quad \text{and} \quad \mathcal{R}_{\mu^{\ast}\nu^{\ast}}=0.
\end{equation}

The conditions $b=a$ and $b=-a$ correspond to the projection of the Lagrangian density (or Ricci tensor) given in (\ref{lagran1}) on the spaces $\mathcal{M}$ and $\mathcal{M}^{\ast}$ respectively. In fact, using the projection tensor fields (\ref{proj}), we easily check that
\begin{equation}
\mathcal{R}_{ik}\mathcal{P}^{k}_{j}= \left(\mathcal{R}_{\mu\nu}, 0\right) \quad \text{and} \quad \mathcal{R}_{ik}\mathcal{Q}^{k}_{j}=\left(0, \mathcal{R}_{\mu^{\ast}\nu^{\ast}} \right).
\end{equation}

{\it''Projections''} on the separate spaces can be taken as a fundamental symmetry in this model. Then, the invariance of the {\it ''bi-field equations''} under this symmetry, always provides one of the cosmological terms to vanish. In other word one of the spaces is not sensitive to the larger vacuum energy \cite{demir2,nima}.

Tackling the problem of the cosmological constant by proposing two universes is not new. It was shown that in a two interacting universes, the so called antipodal symmetry leads to vanishing effective cosmological constant in both universes rather than in an only one \cite{linde1}.

Although it leads to zero cosmological constant, the mechanism proposed in this work can not be considered as solution to the cosmological constant problem. In fact, getting rid of one the cosmological terms automatically creates another in the other space. However, the two spaces may describe  a possible initial and final states of the universe, where a larger vacuum energy in the initial state tends to be zero in later stages. We discuss this possibility later after reformulating the model in the presence of matter fields.

The formalism proposed here can be generalized to the case of matter fields. However, unlike general relativity, coupling general matter fields to gravity in affine theory is not trivial. Certain affine Lagrangian densities including matter have been derived using a covariant Legendre transformations which lead to metric-affine theory from a purely affine theory \cite{kijowski1,kijowski2}.

Now, we take for simplicity a matter source as a simple scalar field $\phi \left( x^{i}\right)$, where $i= 1,...,8$. The affine Lagrangian density is then written as
\begin{equation}
\label{lagran2}
\mathcal{L}=2V^{-1}\left( \phi \right)\sqrt{\text{Det}\left[ \mathcal{R}_{ij}-\partial_{i}\phi \partial_{j}\phi \right] },
\end{equation}
where $V\left( \phi \right)$ is a potential energy. Here and in all what follows, we take the gravitational constant $8\pi G=1$.
\newline
The associated canonical momentum (\ref{pi}) reads
\begin{equation}
\label{pi2}
\pi^{ij}=V^{-1}\left( \phi \right) \sqrt{\text{Det}\left[ \mathcal{R}_{ij}-\partial_{i}\phi \partial_{j}\phi \right] } \left( \kappa^{-1}\mathcal{R}-\partial\phi.\partial\phi\right)^{ij} ,
\end{equation}
which again satisfies the dynamical equation (\ref{dyn}) obtained from the Euler-Lagrange equations (\ref{euler}) for this case.
\newline
From the previous discussion, the resulted field equations with matter in the eight dimensional space become
\begin{equation}
\mathcal{R}_{ij}=\left( a\mathcal{G}_{ij}+b\mathcal{F}_{ij} \right)V\left( \phi\right) + \partial_{i}\phi \partial_{j}\phi.
\end{equation}
These equations again split into two field equations in four dimensional spaces as follows
\begin{eqnarray}
\label{matter1}
\mathcal{R}_{\mu\nu}=\left( a+b\right)V\left( \phi\right)g_{\mu\nu} + \partial_{\mu}\phi \partial_{\nu}\phi,
\end{eqnarray}
\begin{eqnarray}
\label{matter2}
\mathcal{R}_{\mu^{\ast}\nu^{\ast}}=\left( a-b\right)V\left( \phi\right)g_{\mu^{\ast}\nu^{\ast}} + \partial_{\mu^{\ast}}\phi \partial_{\nu^{\ast}}\phi.
\end{eqnarray}
Unlike the vacuum case which is described by the Ricci tensors (\ref{field2}) and (\ref{field3}), the scalar field appears in both universes as we see from the field equations (\ref{matter1}) and (\ref{matter2}). The two universes seem to be connected due to matter fields. The only case that makes them totally disconnected is the case of a strictly constant potential $V_{0}$ (a cosmological constant), this is almost equivalent to the vacuum case described earlier.

In addition to the gravitational field equation (\ref{matter1}) and (\ref{matter2}), the scalar field $\phi$ also satisfies its dynamical equations in both spaces. In fact, taking the Lagrangian density (\ref{lagran2}) and using the form of the scalar density $\pi^{ij}$ given in
(\ref{sol1}) and (\ref{pi2}), the Euler-Lagrange equations for the scalar field $\phi$ in the two spaces read
\begin{eqnarray}
\Box_{g} \phi-\left( a+b\right)V^{\prime}\left( \phi\right)=0 \\
\Box_{g^{\ast}}\phi-\left( a-b\right)V^{\prime}\left( \phi\right)=0
\end{eqnarray}
where $\Box_{g}$ and $\Box_{g^{\ast}}$ are the d'Alembert operators in both spaces $\mathcal{M}$ and $\mathcal{M^{\ast}}$ respectively.

The cosmological evolution of this scalar field (the inflaton) is given by the Friedman equations in the separate spaces. For this, we propose the flat Friedman-Robertson-Walker metric in both spaces $\mathcal{M}$ and $\mathcal{M}^{\ast}$ as follows
\begin{eqnarray}
d s^{2}=-dt^{2}+a^{2}\left( t\right) d \overrightarrow{x}^{2} \\
d s_{\ast}^{2}=-d t_{\ast}^{2}+a^{2}_{\ast} \left( t_{\ast}\right)
d \overrightarrow{x_{\ast}}^{2}
\end{eqnarray}
where the asterisks ($\ast$) always referred to the coordinates in the space $\mathcal{M}^{\ast}$.

These lead to the Friedman equations
\begin{eqnarray}
\label{fr1}
\frac{\overset{\cdot\cdot}{a}}{a}=-\frac{4\pi G}{3}
\left[ 2\dot{\phi}^{2}-2\left(a+b \right)V\left(\phi \right)\right],
\\
\label{fr2}
\frac{\overset{\cdot\cdot}{a_{\ast}}}{a_{\ast}}=-\frac{4\pi G}{3}
\left[ 2\dot{\phi}^{2}-2\left(a-b \right)V\left(\phi \right)\right],
\end{eqnarray}
where the time derivative in the last equation is with respect to $t_{\ast}$.

In the very early universe, it is the potential energy which is quite important than kinetic energy and one may take $\partial_{\mu}\phi \sim 0$. In this case, equation (\ref{matter1}) becomes
\begin{eqnarray}
\label{initial}
\mathcal{R}_{\mu\nu} \sim 2a V\left( \phi\right)g_{\mu\nu},
\end{eqnarray}
where we took the case $b=a$, discussed above (projective symmetry). For this case, equation (\ref{matter2}) automatically reads
\begin{eqnarray}
\label{final}
\mathcal{R}_{\mu^{\ast}\nu^{\ast}}= \partial_{\mu^{\ast}}\phi \partial_{\nu^{\ast}}\phi.
\end{eqnarray}
The field equations (\ref{initial}) and (\ref{final}) (also equivalent to the case $b=-a$) may describe both initial and final states of the universe (before and after inflation) respectively. In fact, inflation is driven by a large potential energy $2aV\left(\phi \right)$ restored in a scalar field. This potential energy which is equivalent to vacuum energy released to kinetic energy and a possible very tiny cosmological constant describing dark energy in the later stage of the universe.

The Friedman equations (\ref{fr1}) and (\ref{fr2}) of these initial and final states of the universe are written as
\begin{eqnarray}
\label{frr1}
\frac{\overset{\cdot\cdot}{a}}{a}=-\frac{4\pi G}{3}
\left[ 2\dot{\phi}^{2}-4aV\left(\phi \right)\right],
\end{eqnarray}
\begin{eqnarray}
\frac{\overset{\cdot\cdot}{a_{\ast}}}{a_{\ast}}=-\frac{4\pi G}{3}
\left( 2\dot{\phi}^{2}\right),
\label{frr2}
\end{eqnarray}
where we took the condition $b=a$, and an equivalent equations can be obtained for $b=-a$.
As in the standard cosmology, the early accelerated phase of the universe (the initial state here) is controlled by the \text{\it gravitational} mass density $\rho+3p$, with $\rho$ and $p$ are referred to density and pressure of the inflaton. This quantity appears on the right hand side of equation (\ref{frr1}), and it is calculated from
\begin{equation}
\rho_{grav}=\rho+3p = 2\dot{\phi}^{2}-4aV\left(\phi \right),
\end{equation}
where we used the density and pressure expressions
\begin{equation}
\rho=\frac{1}{2}\dot{\phi}^{2}+2aV\left(\phi \right),\quad
p=\frac{1}{2}\dot{\phi}^{2}-2aV\left(\phi \right),
\end{equation}
which are resulted from the energy momentum tensor of the scalar field (for $b=a$)
\begin{equation}
T_{\mu\nu}=\partial_{\mu}\phi\partial_{\nu}\phi-\frac{1}{2}g_{\mu\nu}\left[\left(\partial\phi\right)^{2}+4aV\left(\phi \right) \right].
\end{equation}
The initial state discussed above is described by a rapid exponential expansion governed by the potential $aV_{0}$ (we took an almost constant potential). The presence of this potential translates the effects of the \text{\it gravitational} mass density $\rho_{grav}$ which rules the gravitational dynamics.

However, this gravitational mass density is no longer active in the final state. In fact, the kinetic term $\dot{\phi}^{2}$ in equation (\ref{frr2}) is simply the \text{\it inertial} mass density $\rho_{iner}$.
\begin{equation}
\dot{\phi}^{2}= \rho_{\ast}+p_{\ast}=\rho_{iner},
\end{equation}
where $\rho_{\ast}$ and $p_{\ast}$ are the density and pressure of the scalar field in the space $\mathcal{M}^{\ast}$
\begin{eqnarray}
\rho_{\ast}=\frac{1}{2}\dot{\phi}^{2}+\left(a-b\right) V\left(\phi \right)
\\
p_{\ast}=\frac{1}{2}\dot{\phi}^{2}-\left(a-b\right) V\left(\phi \right).
\end{eqnarray}

In the final state of the universe which is described by the space $\mathcal{M}^{\ast}$, the evolution of the scalar field is governed by its \text{\it inertial} mass density rather than the \text{\it gravitational} one. This result is a consequence of the condition $b=a$ (equivalently $b=-a$).

To see this clearly, it is interesting to see the evolution of a general cosmological perfect fluid in this model. Before doing this, we would like to propose a general affine variational action which describes the gravitational dynamics governed by a general perfect fluid given by its energy density and pressure.

Suppose that the universe is dominated by matter field given by its mass energy density $\rho$ and pressure $p$ which are related by its equation of state
\begin{equation}
p=\omega \rho.
\end{equation}

We define the affine action as follows
\begin{equation}
\label{pf}
S_{\text{Aff}}=-\int d^{4} x \frac{4}{\left( \rho-p\right)} \sqrt{\text{Det}\left[\mathcal{R}_{ab} \left(\Gamma \right)- \left(\rho +p \right)u_{a}u_{b} \right] },
\end{equation}
where as in standard cosmology, we defined a preferred 4-vector velocities $u_{a}$.  These vectors are required to define the geodesics in the affine spacetime.

The mean quantities that enters the definition of this affine action are then the affine connection that is incorporated in the Ricci tensor, the directional 4-vector velocities and the mass densities $\rho-p$ and $\rho+p$ \cite{azri3}. An interesting remark here, is that Eddington's gravity is a theory of only a cosmological constant. This is clear from  this action when $\omega=-1$.

Variation of this action with respect to the affine connection leads to the dynamical equation
\begin{eqnarray}
\label{pfdyn1}
\nabla_{c}\pi^{ab}=0,
\end{eqnarray}
where we have put
\begin{eqnarray}
\nonumber
\pi^{ab}=-\frac{2}{\rho-p} \sqrt{\text{Det}\left[\mathcal{R}_{ab} \left(\Gamma \right)- \left(\rho +p \right)u_{a}u_{b} \right] }
\\
\times \left[\mathcal{R} \left(\Gamma \right)- \left(\rho +p \right)u.u \right]^{ab},
\end{eqnarray}
such that the term $\left[\mathcal{R} \left(\Gamma \right)- \left(\rho +p \right)u.u \right]^{ab}$ is the inverse of the quantity $\mathcal{R}_{ab} \left(\Gamma \right)- \left(\rho +p \right)u_{a}u_{b}$.

We solve the dynamical equation (\ref{pfdyn1}) by introducing an invertible tensor field $g_{ab}$ such that
\begin{eqnarray}
\nonumber
\sqrt{g} g^{ab}=-\frac{2}{\rho-p} \sqrt{\text{Det}\left[\mathcal{R}_{ab} \left(\Gamma \right)- \left(\rho +p \right)u_{a}u_{b} \right] }
\\
\times \left[\mathcal{R} \left(\Gamma \right)- \left(\rho +p \right)u.u \right]^{ab}.
\label{pfdyn2}
\end{eqnarray}
Then we get $\nabla_{c}g_{ab}=0$, and the connection becomes a Levi-Civita of $g_{ab}$ which plays a role of a metric tensor.

Now, equation (\ref{pfdyn2}) is equivalent to
\begin{equation}
\label{pff1}
\mathcal{R}_{ab}= \left(\rho+p \right)u_{a}u_{b}-\frac{\left(\rho-p \right)}{2}g_{ab},
\end{equation}
and in terms of Einstein tensor
\begin{equation}
\label{pff2}
\mathcal{R}_{ab}-\frac{1}{2}g_{ab}\mathcal{R}= \left(\rho+p \right)u_{a}u_{b}-pg_{ab}.
\end{equation}

To obtain the last equation, we contracted equation (\ref{pff1}) using $g^{ab}$ and used $g^{ab}u_{a}u_{b}=1$, where the signature of $g$ is supposed to be $\left(+, -, -, -\right)$.

Let us now return to our mean model where the global 8-dim space is taken separable, i.e, a product of two spaces. Without repeating the calculations, one may define the affine action (\ref{pf}) in this global space as we have done in equation (\ref{lagran2}) for the scalar field. In this case, one may easily check that the field equations in the separate factor spaces are given by
\begin{eqnarray}
\label{ed1}
\mathcal{R}_{\mu\nu}= \left(\rho+p \right)u_{\mu}u_{\nu}-\frac{\left( a+b \right) \left(\rho-p \right)}{2} g_{\mu\nu}
\end{eqnarray}
\begin{eqnarray}
\label{ed2}
\mathcal{R}_{\mu^{\ast}\nu^{\ast}}= \left(\rho+p \right)u_{\mu^{\ast}}u_{\nu^{\ast}}-\frac{\left( a-b \right)\left(\rho-p \right)}{2} g_{\mu^{\ast}\nu^{\ast}}.
\end{eqnarray}

In the vacuum case discussed earlier in this section, we have seen that the condition $b=a$ (equivalently $b=-a$) renders the cosmological constant to zero in the final state. In the presence of matter which is usually described by a perfect fluid, this condition eliminates the effects of the \text{\it gravitational} mass density $\rho_{grav}=\rho+3p$ in one of the factor spaces which is considered to be the final state of the universe. This mass density is inactive in this state and the dynamics of the universe is governed by the \text{\it inertial} mass density of the fluid, $\rho_{iner}=\rho+p$. This inertial mass density is completely zero for vacuum energy, and this is indeed the case studied earlier in this section.

As we have seen so far, we have proposed a mechanism based on product spaces which allows us to eliminate the huge vacuum energy which appears trivially in Einstein's field equations. This can not completely solve the cosmological constant problem where the very tiny observed value should appear in the equations to explain the recent accelerating phase of the universe. This phase is again governed by the gravitational mass density of a very small cosmological term. At that end, this model might be developed again, and rather than taking the exact condition $b=a$, one for example may try to relate the constant $a-b$ (and $a+b$) to the gravitational constant and see possible degravitation of the vacuum energy in this construction \cite{demir1}.

We conclude by noticing that unlike general relativity, where the perfect fluid which appears on right hand side of equation (\ref{pff2}) is usually postulated, but here we have derived it from an affine variational principle (\ref{pf}) where the gravitational and inertial mass densities play an important role in its definition. We leave a detailed study of its cosmology to a near future work \cite{azri3}.

\section{Summary}
Affine theory of gravity described by Eddington's action remains the simplest and the more natural theory that one may use to describe gravitational interaction as in Yang Mills theories. More developments and extensions of this theory may lead to understanding the evolution of the early and the late universe quite differently than in the framework of general relativity.

The essential point of this article was the derivation of the separate Einstein spaces from affine variational principle. Originally, these spaces which are product spaces are studied in Riemannian geometry where a global metric tensor is proposed a priori. As in Eddington's theory, we showed that this metric is generated dynamically. These spaces appear as a product of two maximally symmetric spaces with non independent cosmological constants. We discussed the cosmological constant problem in this model by proposing that the invariance under projection on the separate spaces may be considered as a fundamental symmetry to render the cosmological constant of one of the spaces to zero. In the presence of a scalar field as matter field, we showed that these spaces which we name ''separate Einstein-Eddington'' may lead to a description of the initial and final stages of the universe where the initial state is dominated by a large potential energy which tends to zero later.

Other interesting step made in this paper is the affine action principle of a general perfect fluid. This action coincides with Eddington's one in the case of the cosmological constant. This may need more details which we leave for another future work.

Finally, we should mention that this model of ''separate Einstein-Eddington'' space does not elucidate direct solutions to the problems that faced inflationary cosmology, such as the nature of the inflaton or the graceful exit problem \cite{linde2,steinhardt}. However, one may develop this model more and study in details the possible connection of the factor spaces. In fact, finding a mechanism that describes the connection between the two spaces may lead to some possible understanding of some of the problems that we mentioned.

\section*{Acknowledgements}
I am thankful to Durmu{\c s} Ali Demir for suggestions and help. I would like to thank Jerzy Kijowski for helpful Email exchanges, and Recai Erdem for drawing my attention to certain references. I also thank B. Korutlu, C. Karahan, O. Yilmaz and K. Gultekin for some discussions.

\end{document}